# *Ab initio* lattice thermal conductivity of MgSiO$_3$ across perovskite-postperovskite phase transition

Zhen Zhang[1], Renata M. Wentzcovitch[1,2,3,*]

[1]Department of Applied Physics and Applied Mathematics, Columbia University, New York, NY 10027, USA.

[2]Department of Earth and Environmental Sciences, Columbia University, New York, NY 10027, USA.

[3]Lamont–Doherty Earth Observatory, Columbia University, Palisades, NY 10964, USA.

[*]To whom correspondence should be addressed.
rmw2150@columbia.edu

**Abstract**

Lattice thermal conductivity ($\kappa_{lat}$) of MgSiO$_3$ postperovskite (MgPPv) under the Earth's lower mantle high pressure-temperature conditions is studied using the phonon quasiparticle approach by combing *ab initio* molecular dynamics and lattice dynamics simulations. Phonon lifetimes are extracted from the phonon quasiparticle calculations, and the phonon group velocities are computed from the anharmonic phonon dispersions, which in principle capture full anharmonicity. It is found that throughout the lowermost mantle, including the D" region, $\kappa_{lat}$ of MgPPv is ~25% larger than that of MgSiO$_3$ perovskite (MgPv), mainly due to MgPPv's higher phonon velocities. Such a difference in phonon velocities between the two phases originates in the MgPPv's relatively smaller primitive cell. Systematic results of temperature and pressure dependences of both MgPPv's and MgPv's $\kappa_{lat}$ are demonstrated.

# I. INTRODUCTION

The Earth's lower mantle (LM) extends from 670 to 2890 km in depth, corresponding to 55 vol% of the Earth's whole interior. The core-mantle boundary (CMB) located at 2890 km depth is the interface between the rocky LM and the molten metallic core. Upwelling heat is transported by thermal conduction across the CMB where mass transport is impeded [1]. Lattice thermal conductivity ($\kappa_{lat}$) of LM minerals is fundamental to determine CMB heat flow [2] that controls thermal histories of the mantle and core. It is also critical for understanding the dominance of thermal conduction and thermal convection above the CMB. $MgSiO_3$ perovskite (MgPv), the most abundant mineral in the LM, undergoes a phase transition to $MgSiO_3$ postperovskite (MgPPv) above 125 GPa and 2500 K [3–5]. Therefore, MgPPv is believed to be the most abundant mineral in the lowermost mantle, the D" region. With a stacked $SiO_6$ octahedral layered structure, MgPPv appears anisotropic and is believed to be a reasonable explanation for the D" seismic discontinuity with wide topography [3–5]. Investigation of the thermal conductive and thermoelastic properties of MgPPv is essential for understanding the dynamics in the Earth's deep interior.

Reports on $\kappa_{lat}$ of MgPPv are scarce. Under the LM conditions, i.e., $23 < P < 135$ GPa and $2000 < T < 4000$ K [6,7], experimental measurements encounter great difficulties. Ohta *et al.* 2012 [8] conducted measurements of pure MgPPv's $\kappa_{lat}$ up to 141 GPa at room temperature. Recently, Okuda *et al.* [9] conducted measurements for iron-bearing MgPPv at temperatures up to 1560 K. However, grain size effects and iron impurity effects [8,9] introduce large uncertainties in terms of constraining MgPPv's $\kappa_{lat}$ in the lowermost mantle. *Ab initio* calculations provide an alternative to investigate $\kappa_{lat}$ of crystalline minerals at such extreme conditions. Dekura and Tsuchiya [10] calculated $\kappa_{lat}$ of MgPPv up to the CMB conditions by solving the linearized phonon Boltzmann transport equation. However, the perturbative approach adopted by Dekura and Tsuchiya [10] only accounts for up to third-order phonon-phonon scattering processes. Thus, the obtained phonon lifetimes may be overestimated at high temperatures.

In this study, we report the temperature and pressure dependence of MgPPv's $\kappa_{lat}$ up to the CMB conditions using the phonon quasiparticle approach [11,12], a hybrid approach combining *ab initio* molecular dynamics (MD) and lattice dynamics simulations. This approach can address the intrinsic anharmonicity of crystalline materials and has been successfully applied to study anharmonic phonon dispersions [11–14], anharmonic free energies [11,12], phase transition [13], and lattice thermal conductivities [15,16] at high pressure-temperature (*P-T*) conditions. The present methodology to compute $\kappa_{lat}$ on the basis of the phonon quasiparticle approach can be summarized as the following five steps. First, atomic velocities obtained from the *ab initio* MD simulations are projected onto the harmonic phonon polarization vectors obtained using density functional perturbation theory (DFPT) [17] to compute the mode-projected velocity autocorrelation functions (VAF). Second, phonon quasiparticle properties, i.e., renormalized phonon frequencies and phonon lifetimes, are extracted from the VAF. Third, by constructing the effective harmonic dynamical matrices from the renormalized frequencies, anharmonic phonon dispersion throughout the Brillouin zone (BZ) can be obtained via Fourier interpolation. Fourth, phonon lifetimes on dense, converged **q**-meshes are evaluated by relying on the frequency dependence of phonon linewidth, while phonon group velocities at any **q**-point in the BZ can be calculated by calculating **q**-gradients of the anharmonic phonon dispersion. Fifth, with phonon lifetimes and velocities obtained, $\kappa_{lat}$ is determined within the Peierls-Boltzmann theory [18,19] that describes heat carried by phonon quasiparticles.

In principle, the phonon quasiparticle approach captures anharmonic interactions to infinite orders. Therefore, compared to the conventional perturbative methods [10,20,21] accounting for up to three-phonon scatterings, the present approach should yield more reliable phonon lifetimes at high temperatures where higher-order phonon-phonon scatterings arise. Besides, by computing anharmonic phonon dispersion, the obtained phonon velocities are also temperature-dependent. Such features are not present in the conventional perturbative methods to evaluate $\kappa_{lat}$ [10,20,21]. Several other methods can investigate the anharmonic effects of solids non-perturbatively [22–25]. The resulting phonon frequencies and lifetimes can further be used to study $\kappa_{lat}$. Together with our phonon quasiparticle approach, these methods are expected to describe lattice thermal conduction more appropriately beyond the lowest-order three-phonon scattering processes.

## II. METHOD

Thermal conduction in nonmetallic crystalline materials is dominated by phonon-phonon scatterings. Within the relaxation-time approximation (RTA) for the Boltzmann transport equation [18,19], the lattice thermal conductivity can be calculated as

$$\kappa_{lat} = \frac{1}{3}\sum_{\mathbf{q}s} c_{\mathbf{q}s} v_{\mathbf{q}s} l_{\mathbf{q}s}, \qquad (1)$$

where $c_{\mathbf{q}s}$, $v_{\mathbf{q}s}$ and $l_{\mathbf{q}s}$ are phonon heat capacity, group velocity, and mean free path, respectively, of phonon mode (**q**, $s$) with frequency $\omega_{\mathbf{q}s}$. **q** is the phonon wave vector, and $s$ indexes the 3$n$ phonon branches of an *n*-atom primitive cell. The mean free path can further be expressed as $l_{\mathbf{q}s} = v_{\mathbf{q}s}\tau_{\mathbf{q}s}$ where $\tau_{\mathbf{q}s}$ is the phonon lifetime.

In the present phonon quasiparticle approach, a phonon quasiparticle of normal mode (**q**, *s*) is numerically defined by the mode-projected velocity autocorrelation function (VAF),

$$\langle V_{\mathbf{q}s}(0) \cdot V_{\mathbf{q}s}(t) \rangle = \lim_{\tau \to \infty} \frac{1}{\tau} \int_0^\tau V_{\mathbf{q}s}^*(t') V_{\mathbf{q}s}(t'+t) dt', \qquad (2)$$

where $V_{\mathbf{q}s}(t) = \sum_{i=1}^N V(t) \cdot e^{i\mathbf{q}\cdot\mathbf{r}_i} \cdot \hat{\mathbf{e}}_{\mathbf{q}s}$ is the polarized velocity projected onto the (**q**, *s*)-mode harmonic phonon polarization vector, $\hat{\mathbf{e}}_{\mathbf{q}s}$. $\hat{\mathbf{e}}_{\mathbf{q}s}$ is determined by the DFPT [17]. $V(t) = V\left(\sqrt{M_1}\mathbf{v}_1(t), \ldots, \sqrt{M_N}\mathbf{v}_N(t)\right)$ is the mass-weighted velocity with $3N$ components, where $\mathbf{v}_i(t)$ $(i = 1, \ldots, N)$ is the atomic velocity simulated by *ab initio* MD of an $N$-atom supercell, and $M_i$ is the atomic mass of the $i^{th}$ atom in the supercell. For a well-defined phonon quasiparticle, its power spectrum,

$$G_{\mathbf{q}s}(\omega) = \left| \int_0^\infty \langle V_{\mathbf{q}s}(0) \cdot V_{\mathbf{q}s}(t) \rangle e^{i\omega t} dt \right|^2 \qquad (3)$$

has a Lorentzian line shape with a single peak at $\widetilde{\omega}_{\mathbf{q}s}$ and a linewidth of $1/(2\tau_{\mathbf{q}s})$ [12,26], where $\widetilde{\omega}_{\mathbf{q}s}$ is the renormalized frequency, and $\tau_{\mathbf{q}s}$ is the phonon quasiparticle lifetime of mode (**q**, *s*). As reported by previous studies, for a well-defined phonon quasiparticle, converged quasiparticle properties, i.e., $\widetilde{\omega}_{\mathbf{q}s}$ and $\tau_{\mathbf{q}s}$, can be phenomenologically extracted by fitting the VAF to the expression [26],

$$\langle V_{\mathbf{q}s}(0) \cdot V_{\mathbf{q}s}(t) \rangle = A_{\mathbf{q}s} \cos(\widetilde{\omega}_{\mathbf{q}s} t) e^{-t/(2\tau_{\mathbf{q}s})}, \qquad (4)$$

where $A_{\mathbf{q}s}$ is the oscillation amplitude. The obtained $\tau_{\mathbf{q}s}$ can be used to evaluate the lattice thermal conductivity via Eq. (1) [15].

$\widetilde{\omega}_{\mathbf{q}s}$ are further used to produce anharmonic phonon dispersion by constructing the effective harmonic dynamical matrix [11–14],

$$\widetilde{D}(\mathbf{q}) = [\hat{\mathbf{e}}_{\mathbf{q}}] \Omega_{\mathbf{q}} [\hat{\mathbf{e}}_{\mathbf{q}}]^\dagger, \qquad (5)$$

where the diagonal matrix $\Omega_{\mathbf{q}} = \text{diag}[\widetilde{\omega}_{\mathbf{q}1}^2, \widetilde{\omega}_{\mathbf{q}2}^2, \ldots, \widetilde{\omega}_{\mathbf{q}3N}^2]$ contains $\widetilde{\omega}_{\mathbf{q}s}^2$ in the diagonal, and $[\hat{\mathbf{e}}_{\mathbf{q}}] = [\hat{\mathbf{e}}_{\mathbf{q}1}, \hat{\mathbf{e}}_{\mathbf{q}2}, \ldots, \hat{\mathbf{e}}_{\mathbf{q}3N}]$ is the matrix containing harmonic polarization vectors. The effective harmonic force constant matrix, $\widetilde{\Phi}(\mathbf{r})$, is obtained by Fourier transforming $\widetilde{D}(\mathbf{q})$, where the anharmonic interaction is effectively captured. Therefore, $\widetilde{\omega}_{\mathbf{q}'s}$ at any wave vector $\mathbf{q}'$ in the BZ can be obtained by diagonalizing

$$\widetilde{D}(\mathbf{q}') = \sum_{\mathbf{r}} \widetilde{\Phi}(\mathbf{r}) \cdot e^{-i\mathbf{q}'\cdot\mathbf{r}}, \tag{6}$$

from which the temperature-dependent anharmonic phonon dispersions and phonon group velocities, $v_{\mathbf{q}s} = d\widetilde{\omega}_{\mathbf{q}s}/d\mathbf{q}$, can be computed.

We conducted *ab initio* MD simulations using density-functional theory-based Vienna *ab initio* simulation package (VASP) [27] employing the projected-augmented wave method (PAW) [28]. The local density approximation (LDA) was chosen for the electron exchange-correlation functional. MgPPv was simulated with $3 \times 3 \times 2$ (180 atoms) supercells with a Γ **k**-point and a kinetic energy cutoff of 400 eV. MD simulations were carried out in the *NVT* ensemble on a series of volumes (*V*), 69.42, 63.74, 61.58, 59.71 and 58.07 Å$^3$/primitive cell, corresponding to densities ($\rho$), 4.80, 5.23, 5.41, 5.58, and 5.74 g/cm$^3$, respectively, and a series of temperatures between 300 and 4000 K controlled by Nosé thermostat [29]. Each MD simulation ran for over 50 ps with a 1 fs time step. Harmonic phonon polarization vectors were computed by the DFPT [17]. Throughout the *V, T* range investigated in this study, phonon quasiparticles were confirmed to be well-defined, and Eq. (4) was applied to extract the quasiparticle properties. Similar simulations were also carried out for MgPv with $2 \times 2 \times 2$ (160 atoms) supercells [12,15].

### III. RESULTS AND DISCUSSION

For a well-defined VAF, reliable $\widetilde{\omega}_{\mathbf{q}s}$, $\tau_{\mathbf{q}s}$ and $A_{\mathbf{q}s}$ can be obtained by least-square fitting as in Eq. (4), using the first several oscillation periods [12,14], i.e., until the oscillation amplitude decays to its half maximum. Figure 1 showcases the VAF of an acoustic phonon mode of MgPPv at 2000 K and the corresponding fitting curve. The changes in $\widetilde{\omega}_{\mathbf{q}s}$ and $\tau_{\mathbf{q}s}$ by conducting the fitting for several more oscillation periods of the VAF are negligible, indicating the convergence of quasiparticle properties.

Figure 2(a) shows the phonon linewidth ($1/(2\tau_{\mathbf{q}s})$) versus $\widetilde{\omega}_{\mathbf{q}s}$ of MgPPv at different temperatures directly extracted from the phonon quasiparticles via Eq. (4). $\tau_{\mathbf{q}s}$ depends inversely on temperature, caused by the intensification of phonon-phonon scatterings as temperature increases. MgPPv's $\tau_{\mathbf{q}s}$ are then compared with those of MgPv at the same *P-T* conditions in Fig. 2(b). In general, MgPPv's $\tau_{\mathbf{q}s}$ are comparable to MgPv's $\tau_{\mathbf{q}s}$. For the low-frequency acoustic modes, MgPPv's $\tau_{\mathbf{q}s}$ are shorter, while for the rest of high-frequency modes, MgPPv's $\tau_{\mathbf{q}s}$ are slightly longer. Temperature dependences of anharmonic phonon dispersions of MgPPv and MgPv at the same *P-T* are shown in Figs. 3(a) and 3(b), respectively. Like MgPv [12,15], MgPPv's phonon dispersions are weakly temperature-dependent at constant volume; thus, MgPPv is also weakly anharmonic. This comparable degree of intrinsic anharmonicity is reflected in the similarity between MgPPv's and MgPv's $\tau_{\mathbf{q}s}$.

To obtain converged lattice thermal conductivity in the thermodynamic limit ($N \rightarrow \infty$) using Eq. (1), $\tau_{\mathbf{q}s}$ and $v_{\mathbf{q}s}$ are required to be evaluated on much denser **q**-meshes throughout the BZ. Here we rely on the frequency dependence of phonon linewidth [15,16,30,31] to interpolate $\tau_{\mathbf{q}s}$ in the BZ. Our previous studies show that both the quadratic relation, $\frac{1}{\tau} \sim \omega^2$, and cubic relation, $\frac{1}{\tau} \sim \omega^3$, work for MgPv [15,16], yielding $\kappa_{lat}$ slightly lower and higher, respectively, than results obtained using the perturbative method with the same **q**-mesh [21]. Here, MgPPv's phonon linewidths phenomenologically follow very well the quadratic relation. The interpolation was carried out for low-frequency and high-frequency modes separately, guided by the goodness of the quadratic fitting, for both phases. MgPPv's $\tau_{\mathbf{q}s}$ were then obtained with a dense, $10 \times 10 \times 10$ **q**-mesh, and are shown in Fig. 4(a). MgPPv's $v_{\mathbf{q}s}$ were also obtained with the same **q**-mesh from the anharmonic phonon dispersion, and are shown in Fig. 4(b). MgPv's $\tau_{\mathbf{q}s}$ and $v_{\mathbf{q}s}$ obtained with a comparably dense, $8 \times 8 \times 8$ **q**-mesh [15,21] at the same *P-T* are also shown for comparison. At the *P-T* conditions considered here, MgPPv's average phonon lifetime, $\bar{\tau}$, is merely 10% longer than that of MgPv. As mentioned before, in general, MgPPv's shorter $\tau_{\mathbf{q}s}$ for low-frequency acoustic modes and slightly longer $\tau_{\mathbf{q}s}$ for high-frequency modes produces a $\bar{\tau}$ comparable to that of MgPv.

As for the average phonon velocity, MgPPv's $\bar{v}$ is 22% larger than that of MgPv. This is caused by the difference in the primitive cell size, and thus, the difference in the number of phonon branches (30 for MgPPv and 60 for MgPv). The base-centered orthorhombic MgPPv (space group *Cmcm*) is an anisotropic layered structure [3–5], and can be described by a monoclinic primitive cell with two formula units (*Z* = 2) [32]. Figure 3 compares MgPPv's phonon dispersions with those of orthorhombic MgPv (space group *Pbnm*) in their respective BZ. It is visible that the monoclinic MgPPv phase has fewer phonon reflections at the BZ boundary. Due to MgPv's larger primitive cell, its phonon dispersions are folded into a smaller BZ and show less dispersive phonon branches with lower group velocities. This effect leads to MgPPv's $v_{\mathbf{q}s}$ being generally larger than those of MgPv. The resulting MgPPv's $\kappa_{lat}$ is 18% higher than that of MgPv at the *P-T* conditions considered here. Although the 10% enhancement in $\bar{\tau}$ and the 22% enhancement in $\bar{v}$ associated with MgPv to MgPPv phase transition are in good agreement with those reported by Dekura and Tsuchiya [10], the enhancement in $\kappa_{lat}$ is not as large as the 50% reported by them [10]. This is because, in the present study, MgPPv's low-frequency acoustic modes with larger phonon velocities are found to have shorter phonon lifetimes than those of MgPv. Whereas in work by Dekura and Tsuchiya [10], MgPPv's low-frequency acoustic modes with larger velocities have very similar lifetimes to those of MgPv (see Supporting Information of Ref. [10]). Such a difference leads to merely a mild enhancement in $\kappa_{lat}$ in our work instead of a drastic one implied by $\bar{v}$ and $\bar{\tau}$.

$\kappa_{lat}$ of MgPPv were computed from $\tau_{\mathbf{q}s}$ and $v_{\mathbf{q}s}$ obtained with the dense **q**-mesh at several densities and temperatures. For the practical geophysical purpose, the density and temperature dependence of $\kappa_{lat}$ can be described as [33,34]

$$\kappa_{lat} = \kappa_{ref} \left(\frac{T_{ref}}{T}\right)^a \left(\frac{\rho}{\rho_{ref}}\right)^g, \qquad (7)$$

where $g$ is [34]

$$g = b \ln\left(\frac{\rho}{\rho_{ref}}\right) + c. \qquad (8)$$

The choice of the reference density ($\rho_{ref}$) and the reference temperature ($T_{ref}$) is not unique but does not change the density and temperature dependence of $\kappa_{lat}$. Here, by choosing $\rho_{ref}$ = 5.41 g/cm$^3$ and $T_{ref}$ = 2000 K, we obtain the fitting parameters $\kappa_{ref}$, $a$, $b$, and $c$ as 9.4 W/m/K, 0.85, -4.33, and 4.93, respectively, for MgPPv. The fitted results are shown as solid curves in Fig. 5(a). At each density, $\kappa_{lat}$ obeys the $1/T^{0.85}$ law, which is milder than the $1/T$ dependence reported by Dekura and Tsuchiya [10]. Nevertheless, within the LM relevant temperature range, i.e., 2000 < $T$ < 4000 K, the $1/T$ relation roughly also applies. The presence of low concentration iron alloying in MgPPv can further flatten the temperature dependence of $\kappa_{lat}$, e.g., $1/T^{0.65}$ as reported by Okuda *et al.* [9].

$\kappa_{lat}(T,\rho)$ are then converted into $\kappa_{lat}(T,P)$ by adopting the well-established quasiharmonic equation of state (EoS) [5,32]. Due to the weak temperature dependence of the phonon dispersions (see Fig. 3(a)), the impact of anharmonic free energy on the high *P-T* EoS is disregarded in this study. The converted $\kappa_{lat}(T,P)$ are shown in Fig. 5(b). At each temperature, $\kappa_{lat}$ depends linearly on pressure, which is a typical feature produced by both experiments and *ab initio* calculations for other major LM phases, e.g., MgPv [8,15,21], MgO [35,36], and cubic CaPv [16]. Previously reported experimental measurements for MgPPv's $\kappa_{lat}$ at room temperature by Ohta *et al.* 2012 [8] are shown for comparison. Within the experimental and computational uncertainties, our predictions agree very well with the measurements. MgPPv's $\kappa_{lat}$ calculated at room temperature using semiclassical nonequilibrium MD simulations [37] are also shown for comparison. Their results were obtained using two different sets of interatomic potentials [37] and agree relatively well with our predictions and the experiments at room temperature. However, with $1/T^{0.5}$ dependence, $\kappa_{lat}$ reported by Ammann *et al.* [37] deviates from our results at high temperatures. The insufficiency of the interatomic potentials could cause this different dependence at high temperatures [10]. Thus, their high temperatures results are not shown for comparison.

Recently reported *ab initio* results for MgPPv's $\kappa_{lat}$ by Dekura and Tsuchiya [10] are significantly larger than ours and experiments. This difference seems to originate in their reported much larger phonon lifetimes [10]. For example, at the same *P-T* conditions, phonon lifetimes reported in the Supporting Information of Ref. [10] are much larger than the results shown in Fig. 4(a), especially for the high-frequency modes. This difference results in their larger $\kappa_{lat}$, 12.0 W/m/K [10], compared to ours, 9.8 W/m/K. The perturbative approach adopted by Dekura and Tsuchiya [10] considers only up to third-order phonon-phonon interactions. The present phonon quasiparticle approach, in principle, captures anharmonic effects to infinite orders. Therefore, accounting for higher-order phonon-phonon scattering processes as done by the phonon quasiparticle approach, as expected, reduces phonon lifetimes.

To gain insights into the effect of the MgPv to MgPPv phase transition on the thermal conductivity of the LM, similar calculations of $\kappa_{lat}$ were also carried out for MgPv [12,15]. The $\kappa_{lat}(T,\rho)$ results fitted to Eq. (7) produced $\rho_{ref}$ = 5.33 g/cm$^3$, $T_{ref}$ = 2000 K, and $\kappa_{ref}$, *a*, *b* and *c* as 7.9 W/m/K, 1.01, -7.85, and 4.37, respectively, for MgPv. The results are shown in Fig. 6(a) for comparison. The converted $\kappa_{lat}(T,P)$ obtained using a quasiharmonic EoS [38] are shown in Fig. 6(b). At each temperature, $\kappa_{lat}(T,P)$ varies linearly with pressure. At room temperature, MgPv's $\kappa_{lat}$ agrees reasonably well with *ab initio* perturbative calculations using the same **q**-mesh reported by Ghaderi *et al.* [21], which further justifies the present methodology. Previously reported experimental measurements at room temperature differ. Manthilake *et al.*'s [33] measurement agrees well with our predictions, while Ohta *et al.*'s [8,39] are lower. Such discrepancy between measurements is likely caused by different grain sizes of MgPv samples used in the measurements [8,33]. As indicated in Refs. [15,21], according to private communications with G. M. Manthilake and K. Ohta, the grain sizes of measured samples by Manthilake *et al.* [33] are 10–15 μm, while those by Ohta *et al.* [8] are ~1 μm. By using samples with larger grain sizes, Manthilake *et al.* [33] find $\kappa_{lat}$ closer to the *ab initio* results in the thermodynamic limit.

With MgPPv's and MgPv's $\kappa_{lat}(T,P)$ obtained, their respective $\kappa_{lat}$ along a typical LM geotherm [40] with a thermal boundary layer above the CMB are obtained and shown in Fig. 7. $\kappa_{lat}$ of MgPPv is only shown in the pressure range investigated in this study, i.e., $P > 40$ GPa. We obtain MgPPv's $\kappa_{lat} = 7.1(4)$ W/m/K at the top of the D" layer at 2600 km depth ($T = 2735$ K and $P = 120$ GPa), and $\kappa_{lat} = 5.9(3)$ W/m/K at the CMB at 2890 km depth ($T = 3625$ K and $P = 135$ GPa). Throughout the D" region, $\kappa_{lat}$ of MgPPv is approximately 25% greater than that of MgPv, which contrasts with the 72% enhancement reported by Ohta *et al.* 2012 [8], and the 50% enhancement reported by Dekura and Tsuchiya [10]. In addition, the temperature dependence of MgPPv's $\kappa_{lat}$ ($1/T^{0.85}$) is milder than that of MgPv ($1/T^{1.01}$). Therefore, in the colder regions of D" layer where MgPPv is expected to be more stable and more abundant [5], the enhancement in $\kappa_{lat}$ due to MgPv to MgPPv phase transition should be even less than 25%. As discussed above, such enhancement at $T = 2000$ K and $P = 116$ GPa is merely 18%.

Next, we estimate the effect of MgPv to MgPPv phase transition on the thermal conductivity of the LM at the CMB. Here we assume a pyrolitic LM composition, i.e., MgPv/MgPPv + MgO with volume proportion 80:20 [10,33]. MgO's $\kappa_{lat} = 35$ W/m/K at the CMB is adopted from Tang and Dong [20]. Then by applying the Hashin-Shtrikman averaging scheme [10,33,41], we obtain $\kappa_{lat}$ of the MgPv + MgO aggregate to be 7.9(3) W/m/K, and $\kappa_{lat}$ of the MgPPv + MgO aggregate to be 9.4(3) W/m/K at the CMB. The MgPv to MgPPv phase transition enhances the $\kappa_{lat}$ of the pyrolitic LM by ~20%, which is less than the ~60% enhancement predicted by Ohta *et al.* 2012 [8], and the ~40% enhancement predicted by Dekura and Tsuchiya [10]. Note here for simplicity, the effects of iron impurities in MgPv, MgPPv and MgO are not accounted for. If the same impurity effects are assumed for MgPv, MgPPv and MgO, i.e., a 50% reduction in $\kappa_{lat}$ [33], then our conclusion of the enhancement in $\kappa_{lat}$ is not changed.

**IV. CONCLUSIONS**

In summary, by combining *ab initio* molecular dynamics and lattice dynamics simulations, we have computed the phonon quasiparticles [11,12] of MgSiO$_3$ postperovskite (MgPPv) under the lower mantle (LM) conditions. Phonon lifetimes and group velocities are extracted from the phonon quasiparticles, and are interpolated in the Brillouin zone to approach the thermodynamic limit ($N \rightarrow \infty$). The obtained phonon lifetimes and velocities are used to evaluate the lattice thermal conductivity ($\kappa_{lat}$) of MgPPv within the relaxation-time approximation for the Boltzmann transport equation. $\kappa_{lat}(T,\rho)$ are converted into $\kappa_{lat}(T,P)$ by using MgPPv's quasiharmonic equation of state [5,32]. It is found that MgPPv's $\kappa_{lat}$ varies as $1/T^{0.85}$ with temperature, and depends linearly on pressure. With $\kappa_{lat}$ along a typical geotherm including a thermal boundary layer, we obtain MgPPv's $\kappa_{lat}$ = 7.1(4) W/m/K at the top of the D" layer, and $\kappa_{lat}$ = 5.9(3) W/m/K at the core-mantle boundary (CMB). Similar calculations were performed for $MgSiO_3$ perovskite (MgPv) and a comparison between MgPv's and MgPPv's $\kappa_{lat}$ is made. It is found that throughout the D" region of the LM, MgPPv's $\kappa_{lat}$ is 25% greater than that of MgPv, which is a smaller difference than the previously reported [8,10]. Such a difference between MgPv's and MgPPv's $\kappa_{lat}$ should be even smaller in cold subducting plates in the D" region, where MgPPv is more stable and abundant [5]. Our calculations also suggest that the MgPv to MgPPv phase transition causes a ~20% enhancement in $\kappa_{lat}$ of the pyrolitic LM aggregate at the CMB.

## ACKNOWLEDGMENTS

This work was primarily funded primarily by the US Department of Energy Grant DE-SC0019759 and in part by the National Science Foundation (NSF) award EAR-1918126. This work used the Extreme Science and Engineering Discovery Environment (XSEDE), USA, which was supported by the NSF Grant ACI-1548562. Computations were performed on Stampede2, the flagship supercomputer at the Texas Advanced Computing Center (TACC), the University of Texas at Austin, generously funded by the NSF through Grant ACI-1134872.

## REFERENCES

[1]  T. Lay, J. Hernlund, and B. A. Buffett, Nat. Geosci. **1**, 25 (2008).
[2]  A. M. Hofmeister, Science **283**, 1699 (1999).
[3]  M. Murakami, K. Hirose, K. Kawamura, N. Sata, and Y. Ohishi, Science **304**, 855 (2004).
[4]  A. R. Oganov and S. Ono, Nature **430**, 445 (2004).
[5]  T. Tsuchiya, J. Tsuchiya, K. Umemoto, and R. M. Wentzcovitch, Earth Planet. Sci. Lett. **224**, 241 (2004).

[6] A. Zerr, X. Diegeler, and R. Boehler, Science **281**, 243 (1998).

[7] R. M. Wentzcovitch, B. B. Karki, M. Cococcioni, and S. de Gironcoli, Phys. Rev. Lett. **92**, 018501 (2004).

[8] K. Ohta, T. Yagi, N. Taketoshi, K. Hirose, T. Komabayashi, T. Baba, Y. Ohishi, and J. Hernlund, Earth Planet. Sci. Lett. **349–350**, 109 (2012).

[9] Y. Okuda, K. Ohta, A. Hasegawa, T. Yagi, K. Hirose, S. I. Kawaguchi, and Y. Ohishi, Earth Planet. Sci. Lett. **547**, 116466 (2020).

[10] H. Dekura and T. Tsuchiya, Geophys. Res. Lett. **46**, 12919 (2019).

[11] T. Sun, D.-B. Zhang, and R. M. Wentzcovitch, Phys. Rev. B **89**, 094109 (2014).

[12] D.-B. Zhang, T. Sun, and R. M. Wentzcovitch, Phys. Rev. Lett. **112**, 058501 (2014).

[13] Y. Lu, T. Sun, Ping Zhang, P. Zhang, D.-B. Zhang, and R. M. Wentzcovitch, Phys. Rev. Lett. **118**, 145702 (2017).

[14] Z. Zhang, D.-B. Zhang, T. Sun, and R. M. Wentzcovitch, Comput. Phys. Commun. **243**, 110 (2019).

[15] D.-B. Zhang, P. B. Allen, T. Sun, and R. M. Wentzcovitch, Phys. Rev. B **96**, 100302(R) (2017).

[16] Z. Zhang, D.-B. Zhang, K. Onga, A. Hasegawa, K. Onta, K. Hirose, and R. M. Wentzcovitch, arXiv:2005.08289.

[17] S. Baroni, S. de Gironcoli, A. D. Corso, and P. Giannozzi, Rev. Mod. Phys. **73**, 515 (2001).

[18] J. M. Ziman, *Electrons and Phonons: The Theory of Transport Phenomena in Solids.* (Oxford Univ. Press, Oxford, 2001).

[19] T. Sun and P. B. Allen, Phys. Rev. B **82**, 224305 (2010).

[20] X. Tang and J. Dong, Proc. Natl. Acad. Sci. U.S.A. **107**, 4539 (2010).

[21] N. Ghaderi, D.-B. Zhang, H. Zhang, J. Xian, R. M. Wentzcovitch, and T. Sun, Sci. Rep. **7**, 5417 (2017).

[22] N. K. Ravichandran and D. Broido, Phys. Rev. B **98**, 085205 (2018).

[23] Y. Xia, K. Pal, J. He, V. Ozoliņš, and C. Wolverton, Phys. Rev. Lett. **124**, 065901 (2020).

[24] O. Hellman and I. A. Abrikosov, Phys. Rev. B **88**, 144301 (2013).

[25] M. Simoncelli, N. Marzari, and F. Mauri, Nat. Phys. **15**, 809 (2019).

[26] T. Sun, X. Shen, and P. B. Allen, Phys. Rev. B **82**, 224304 (2010).

[27] G. Kresse and J. Furthmüller, Phys. Rev. B **54**, 11169 (1996).

[28] P. E. Blöchl, Phys. Rev. B **50**, 17953 (1994).

[29] W. G. Hoover, Phys. Rev. A **31**, 1695 (1985).

[30] P. G. Klemens, Proc. R. Soc. Lond. A **208**, 108 (1951).

[31] C. Herring, Phys. Rev. **95**, 954 (1954).

[32] J. Tsuchiya, T. Tsuchiya, and R. M. Wentzcovitch, J. Geophys. Res. **110**, B02204 (2005).

[33] G. M. Manthilake, N. de Koker, D. J. Frost, and C. A. McCammon, Proc. Natl. Acad. Sci. U.S.A. **108**, 17901 (2011).

[34] S. Imada, K. Ohta, T. Yagi, K. Hirose, H. Yoshida, and H. Nagahara, Geophys. Res. Lett. **41**, 4542 (2014).

[35] D. A. Dalton, W.-P. Hsieh, G. T. Hohensee, D. G. Cahill, and A. F. Goncharov, Sci. Rep. **3**, 2400 (2013).

[36] L. Lindsay, D. A. Broido, J. Carrete, N. Mingo, and T. L. Reinecke, Phys. Rev. B **91**, 121202(R) (2015).

[37] M. W. Ammann, A. M. Walker, S. Stackhouse, J. Wookey, A. M. Forte, J. P. Brodholt, and D. P. Dobson, Earth Planet. Sci. Lett. **390**, 175 (2014).

[38] B. B. Karki, R. M. Wentzcovitch, S. de Gironcoli, and S. Baroni, Phys. Rev. B **62**, 14750 (2000).

[39] K. Ohta, T. Yagi, and K. Hirose, Am. Mineral. **99**, 94 (2014).

[40] F. D. Stacey and P. M. Davis, *Physics of the Earth.* (Cambridge Univ. Press, Cambridge, 2008).

[41] Z. Hashin and S. Shtrikman, Phys. Rev. **130**, 129 (1963).

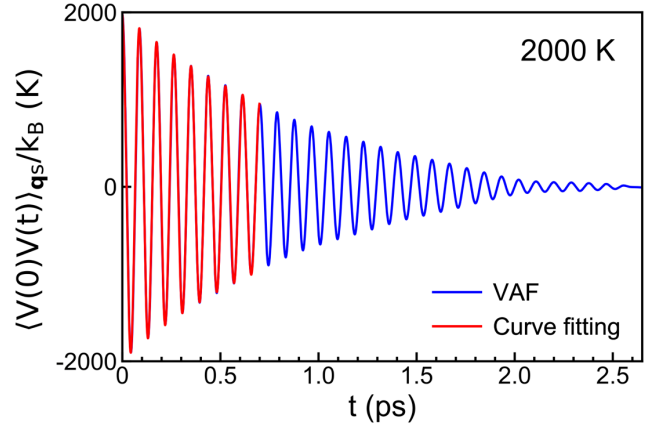

FIG. 1. Mode-projected velocity autocorrelation function (VAF) (blue curve) of an acoustic phonon mode at $\mathbf{q} = (0, \frac{1}{3}, \frac{1}{2})$ with a harmonic frequency of 387 cm$^{-1}$ for MgPPv at 5.41 g/cm$^3$ and 2000 K. The curve fitting (red curve) was conducted for the first few oscillation periods of the VAF until the oscillation amplitude decays to its half maximum.

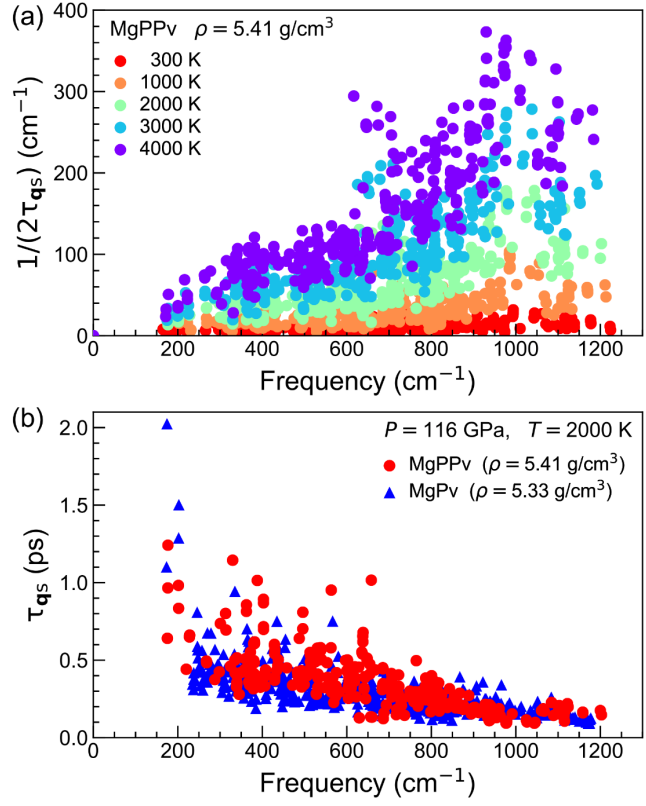

FIG. 2. (a) Phonon linewidth ($1/(2\tau_{\mathbf{q}s})$) versus renormalized phonon frequency ($\widetilde{\omega}_{\mathbf{q}s}$) sampled by MD simulations using a $3 \times 3 \times 2$ supercell (180 atoms) of MgPPv at a series of temperatures. (b) Phonon lifetimes ($\tau_{\mathbf{q}s}$) versus $\widetilde{\omega}_{\mathbf{q}s}$ of MgPPv are compared with those of MgPv using a $2 \times 2 \times 2$ supercell (160 atoms) at $P$ = 116 GPa and $T$ = 2000 K.

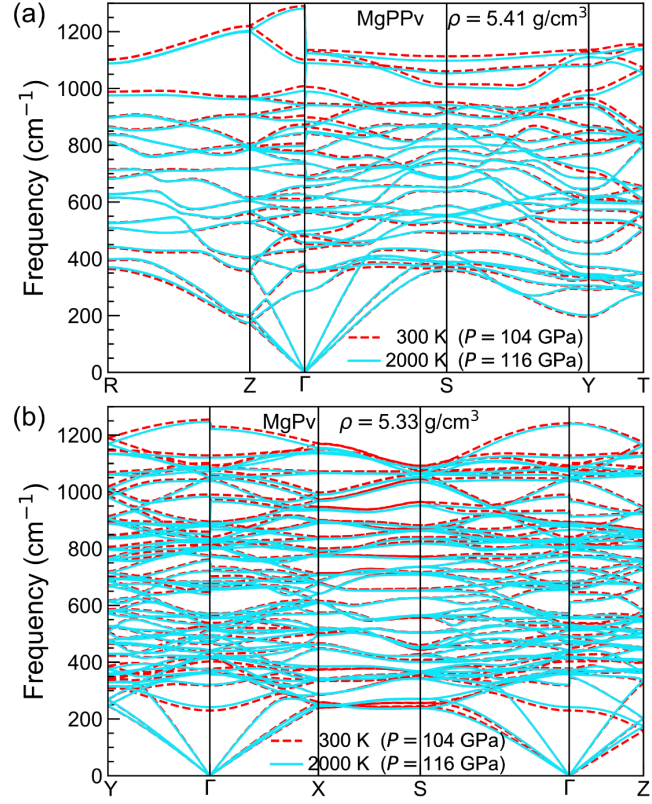

FIG. 3. (a) Anharmonic phonon dispersions at 300 K (dashed red curves) and 2000 K (solid blue curves) at $\rho$ = 5.41 g/cm$^3$ of MgPPv, corresponding to $P$ = 104 and 116 GPa, respectively. (b) Anharmonic phonon dispersions at 300 K (dashed red curves) and 2000 K (solid blue curves) at $\rho$ = 5.33 g/cm$^3$ of MgPv, corresponding to $P$ = 104 and 116 GPa, respectively.

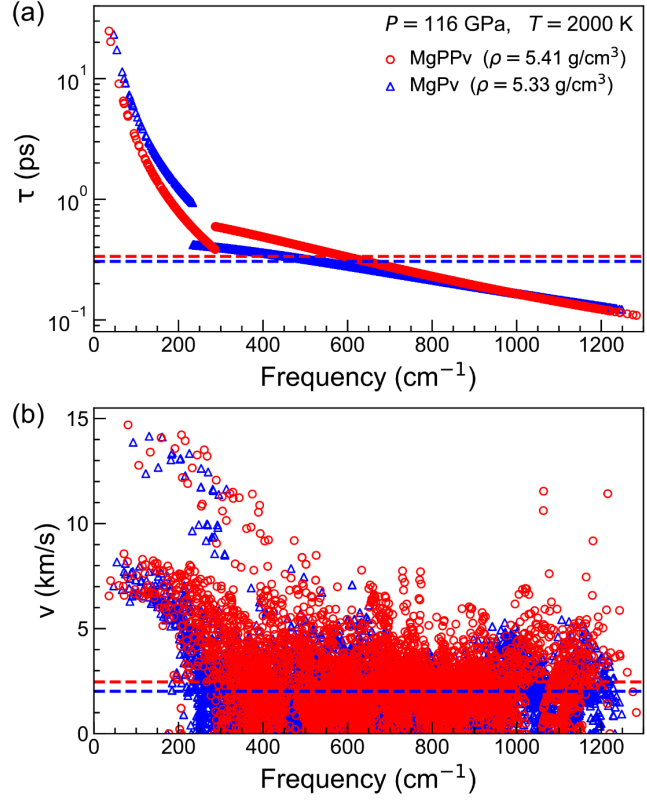

FIG. 4. (a) $\tau_{\mathbf{q}s}$ and (b) $v_{\mathbf{q}s}$ versus $\widetilde{\omega}_{\mathbf{q}s}$ sampled by a $10 \times 10 \times 10$ **q**-mesh of MgPPv at $P = 116$ GPa and $T = 2000$ K are shown as red circles. $\tau_{\mathbf{q}s}$ and $v_{\mathbf{q}s}$ versus $\widetilde{\omega}_{\mathbf{q}s}$ sampled by a $8 \times 8 \times 8$ **q**-mesh of MgPv at the same *P-T* conditions are shown as blue triangles for comparison. Dashed lines indicate average phonon lifetimes ($\bar{\tau}$) and average phonon velocities ($\bar{v}$).

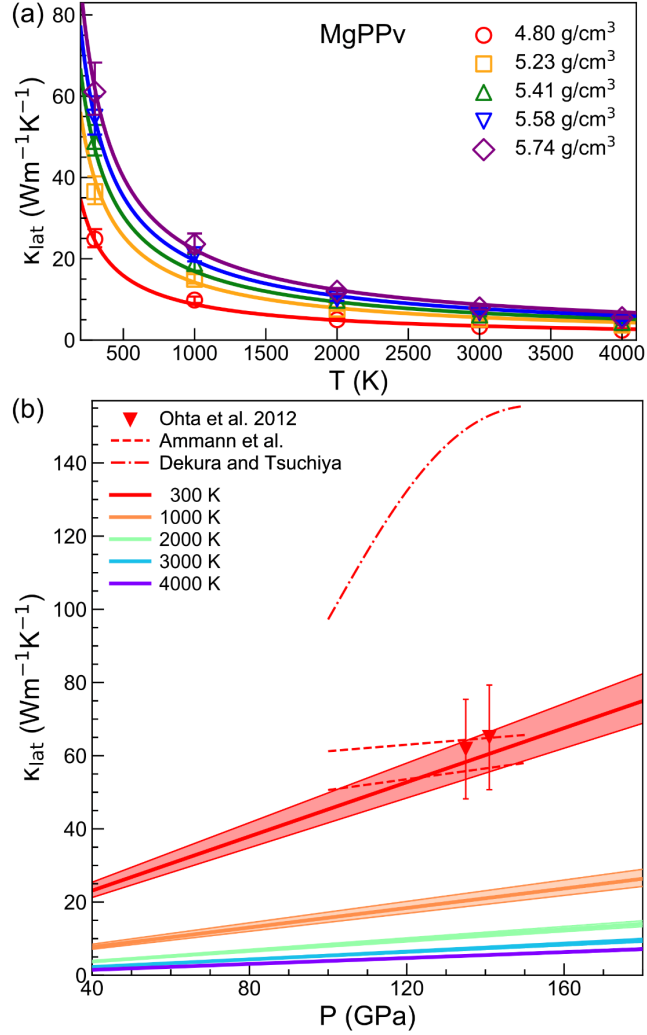

FIG. 5. (a) Lattice thermal conductivity ($\kappa_{lat}$) versus $T$ of MgPPv (symbols) for a series of densities. Error bars show the computational uncertainties by fitting the phonon quasiparticle linewidth versus frequency. (b) $\kappa_{lat}$ versus $P$ of MgPPv at a series of temperatures. Shaded areas indicate the same computational uncertainties as in (a). Symbols with error bars show the previous experimental measurements at 300 K [8]. Dashed lines show the previous semiclassical nonequilibrium MD calculations by adopting two different sets of interatomic potentials at 300 K [37]. Dashed-dotted line shows the previous *ab initio* perturbative calculations at 300 K [10].

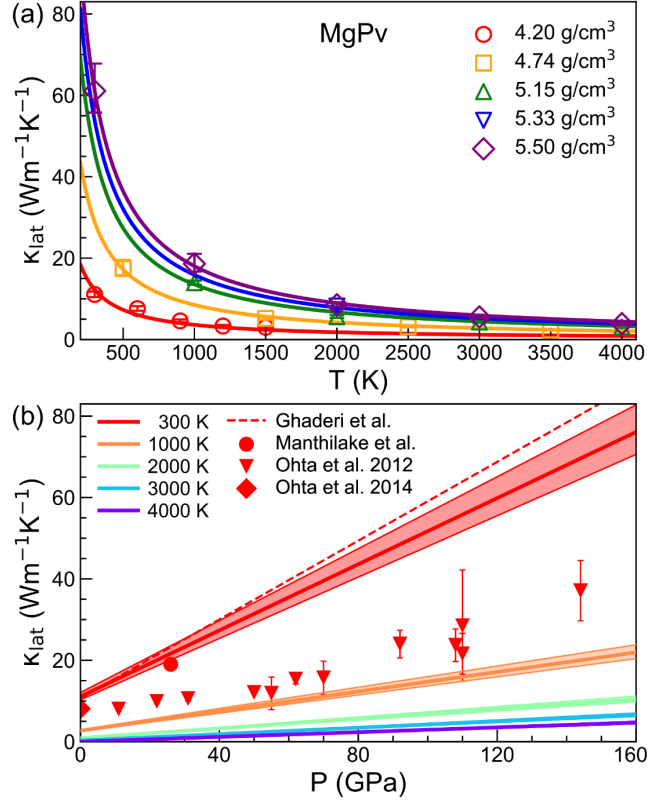

FIG. 6. (a) $\kappa_{lat}$ versus $T$ of MgPv (symbols) for a series of densities. Error bars show the computational uncertainties by fitting the phonon quasiparticle linewidth versus frequency. (b) $\kappa_{lat}$ versus $P$ of MgPv at a series of temperatures. Shaded areas indicate the same computational uncertainties as in (a). Symbols with error bars show previous experimental measurements at 300 K [8,33,39]. Dashed line shows the previous *ab initio* perturbative calculations at 300 K [21].

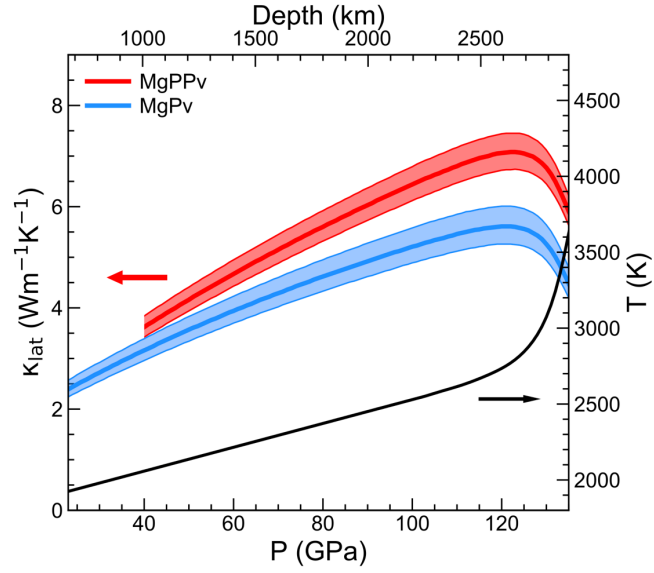

FIG. 7. $\kappa_{lat}$ of MgPPv and MgPv along the geotherm (black curve) [40] are shown in red and blue, respectively. Shaded areas indicate the computational uncertainties. MgPPv's $\kappa_{lat}$ is only shown within the investigated pressure range ($P > 40$ GPa).